\documentclass[reprint,prl,amsmath,amssymb,twocolumn,showpacs]{revtex4-1}

\usepackage{graphicx} 
\usepackage{dcolumn} 
\usepackage{bm} 
\usepackage{amsmath}
\usepackage{braket}
\usepackage{natbib}

\begin{document}


\title{Spin Current Generation and Magnetic Response in Carbon Nanotubes by the Twisting Phonon Mode}


\author{Masato Hamada, Takehito Yokoyama, and Shuichi Murakami}
\affiliation{Department of Physics, Tokyo Institute of Technology, 2-12-1 Ookayama, Meguro-ku, Tokyo 152-8551, Japan}


\date{\today}

\begin{abstract}
We theoretically investigate spin current and magnetic response induced by the twisting phonon mode in carbon nanotubes via the spin-rotation coupling. 
An effective magnetic field due to the twisting mode induces both spin and orbital magnetizations. The induced spin and orbital magnetizations have
both radial and axial components. 
We show that AC pure spin current is generated by the twisting phonon mode. The magnitude of the spin current and orbital magnetization for a (10,10) armchair nanotube is estimated as an example. We find that the AC pure spin current is detectable in magnitude when the frequency of the twisting mode is of the order of GHz, 
and that the orbital magnetization is found to be larger than the spin magnetization. 
\end{abstract}


\maketitle


{\it Introduction}. Carbon nanotubes (CNTs) have been studied extensively, and today various interesting physical characters of CNTs have been reported.
CNTs can be regarded as rolls of graphene sheets, and  therefore their electronic bands
are closely related with those of graphene. 
In graphene, the electronic bands form linear dispersion called a Dirac cone. By rolling the graphene into a CNT, the band structure 
becomes a metal or semiconductor depending on its chirality. 
From the viewpoint of spin transport, 
the spin-orbit interaction of CNT is so weak that spin current cannot be generated for CNT alone by usual methods which require strong spin-orbit interaction.

 In this paper, we propose a new possibility of the spin current generation by using a twisting mode, one of the characteristic phonon modes in 
the CNT \cite{Kane1,TW01,TW02}. 
Our proposal of spin current generation by the twisting mode is based on a spin-rotation coupling. 
The spin-rotation coupling can be derived in relativistic quantum mechanics \cite{Matsuo02, SRC01, SRC02}, and 
it couples mechanical rotation with spins~\cite{Matsuo01,Matsuo02}. 
Physically 
it can be regarded as the Barnett effect \cite{Barnet}; mechanical rotation serves as an 
effective magnetic field, and couples with spin via Zeeman interaction. 
In fact, this effective magnetic field couples with electrons not only via spins but also via orbitals, as seen from Ref.~\cite{Matsuo02}. Thus, 
generation of both spin and orbital magnetizations by the twisting mode is expected. 
%
We show that in CNTs, AC pure spin current is generated by the twisting mode. We also show that the twisting mode induces an orbital magnetization. 
We find that the orbital magnetization is larger than the spin magnetization for realistic parameters. We estimate the magnitude of the spin current and orbital magnetization for a (10,10) armchair nanotube as an example, 
and we find that the AC pure spin current is detectable in magnitude when the frequency of the twisting mode is of the order of GHz. 

Since it is shown that in CNTs and graphene, spins have a relatively long lifetime, they are promising as  leads for spintronics devices. 
Spin current injection into CNT were reported in Refs.~\cite{experiment1, parameter01}, and 
the methods of spin current injection are spin valves and non-local spin injection, both of which need a ferromagnet to inject spin. On the contrary, the generation of spin current by the twisting mode does not require a ferromagnet to generate spin current, and would pave the way for spintronics application of CNTs. 
 
\begin{figure}
\includegraphics[width=7cm]{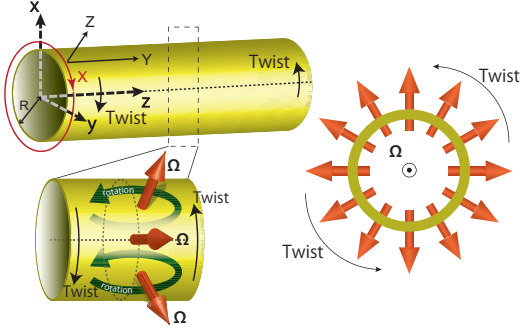}
\caption{(Color online) Schematic figure of the twisting mode in the CNT. $(X,Y,Z)$ and $(x,y,z)$ denote the cylindrical and the Cartesian coordinates, respectively. The angular velocity vector $\mathbf{\Omega}$ due to twisting mode has components along and perpendicular to the tube axis.}
\label{structure}
\end{figure}

{\it Twisting mode}. Among the three types of  phonon modes in CNTs, i.e. stretching, breathing, and twisting modes \cite{Suzuura01}, we focus on the twisting mode, which represents a deformation along the circumference of the tube.
In the Cartesian coordinate $(x,y,z)$ shown in Fig. \ref{structure}, the displacement of the twisting mode is given by 
$u_{\rm t}(z,t)=\theta_0 \sin{(q_{\rm t}z-\omega_{\rm t} t)}$
with the angular amplitude $\theta_0$, wavenumber of the twisting mode $q_{\rm t}$, and frequency of the twisting mode $\omega_{\rm t}= v_{\rm t} q_{\rm t}$. Here $v_{\rm t}$ is the velocity of the mode given by $v_{\rm t} = \sqrt{{\mu}/{M}}$ with shear modulus $\mu$ and the mass of a carbon atom $M$.
We note that the dispersion of the 
twisting mode is almost independent of the chirality of CNT, since the phonon dispersion in a CNT is almost the same as  that in graphene \cite{TW02}.
 The angular velocity is given by
\begin{eqnarray}
\Omega(z,t) = \frac{\partial u_t}{\partial t} = -\theta_0\omega_{\rm t} \cos{(q_{\rm t}z-\omega_{\rm t} t)} \label{z-component} .
\end{eqnarray}
Then 
the angular velocity vector $\mathbf{\Omega}(x,y,z)$, which characterizes
the local rotation of the CNT in Fig. \ref{structure}, is 
\begin{equation}
\mathbf{\Omega} = \frac{1}{2}\left(-\frac{\partial \Omega(z,t)}{\partial z}x,-\frac{\partial \Omega(z,t)}{\partial z}y,2\Omega(z,t) \right), 
\end{equation}
with $x^2+y^2=R^2$, and the tube radius $R$.
This angular velocity works as an effective magnetic field  $\mathbf{B}_{\rm eff} = -(2m_{\rm e}/e) \mathbf{\Omega}$ ~\cite{Matsuo03}
due to spin-rotation coupling, where $-e$ is the electron charge and $m_{\rm e}$ is the electron mass.
Hence in the present case, the effective magnetic field and the corresponding vector potential are given as
\begin{eqnarray}
\mathbf{B}_{\rm eff} &=& -\frac{m_{\rm e}}{e}\left(-\frac{\partial \Omega(z,t)}{\partial z}x,-\frac{\partial \Omega(z,t)}{\partial z}y,2\Omega(z,t)\right), \label{effective magnetic field} \\
\mathbf{A}_{\rm eff} &=& -\frac{m_{\rm e}}{e}(-\Omega(z,t)y,\Omega(z,t)x,0).
\end{eqnarray}
The $z$ component $B_{\rm eff}^{z}$ is along the axial direction, while the other components are along the radial direction.
This effective magnetic field affects both the electronic spins and orbitals. First, it enters the Zeeman coupling:
\begin{align}
H_{\rm s} &= \mu_{\rm B} {\bm \sigma}\cdot \frac{\mathbf{B_{\rm eff}}}{2} = -\frac{\hbar}{2}{\bm \sigma}\cdot \mathbf{\Omega},
\end{align}
where $\mu_{\rm B}= e\hbar/2m_{\rm e}$ is the Bohr magneton. Second, the orbital magnetic field enters the  Hamiltonian via minimal coupling: $\mathbf{p} \rightarrow \mathbf{p} + e\mathbf{A}_{\rm eff}$, which affects electronic orbital motions.

\begin{figure}
\includegraphics[width=8.6cm]{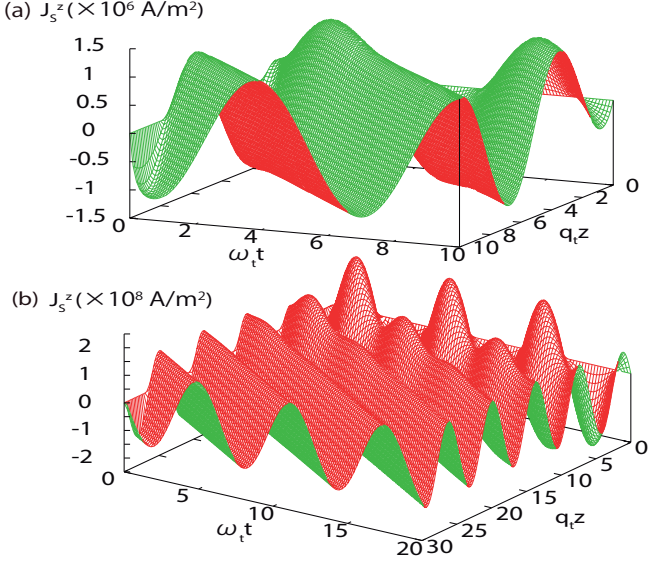}
\caption{\label{spin-current}(Color online) Spin current by the twisting mode in the $(10,10)$ armchair CNT for  (a) $\omega_{\rm t} = 1.2\times 10^{10}\ \mathrm{s^{-1}}$, $q_{\rm t} = 1.2\times 10^{6}\ \mathrm{m^{-1}}$ and (b) $\omega_{\rm t} = 1.2\times 10^{12}\ \mathrm{s^{-1}}$, $q_{\rm t} = 1.2\times 10^{8}\ \mathrm{m^{-1}}$. 
 (a) and (b)  correspond to 
$q_{\rm t}\lambda_{\rm sf}\sim 1$ and 
$q_{\rm t}\lambda_{\rm sf}\sim 100$, respectively.
Near the edge of CNT $(q_{\rm t} z \ll 1)$, spin current is enhanced by the reflection at the edge, and in the region sufficiently away from the edge $(q_{\rm t} z \gg 1)$, spin current behaves similarly to the twisitng mode. }
\end{figure}

{\it Spin current}. We will calculate the spin current in  CNTs induced by the effective 
Zeeman magnetic field due to the twisting mode. We consider the twisting phonon mode in the semi-infinite system $(z\ge0)$ of CNTs. 
We use the spin diffusion equation and calculate spin current polarized along the $z$-axis, using the formalism proposed in Ref.~\cite{Matsuo01}. 
Here we retain only the $z$-component of the Zeeman magnetic field $\mathbf{B}_{\rm eff}$ and neglect the other components, i.e.  
$|{B}_{\rm eff}^z|\gg |{B}_{\rm eff}^x|, |{B}_{\rm eff}^y|$, because the twisting mode is assumed to have a long wavelength. We will discuss justification of this assumption later. 
The spin-rotation coupling generates a difference of chemical potentials between up- and down-spins. The spin diffusion equation  with spin-rotation coupling \cite{Matsuo01} is written as 
\begin{equation}
\left(\frac{\partial}{\partial t} - D\frac{\partial^2}{\partial z^2} + \tau_{\rm sf}^{-1}\right)\delta\mu(z,t) = -\hbar\frac{\partial}{\partial t}\Omega(z,t) \label{spin-diff},
\end{equation}
with the diffusion constant $D$, the spin lifetime $\tau_{\rm sf}$. The spin current polarized along the $z$-axis is then given by 
\begin{equation}
J_{S}^z(z,t) = \frac{\sigma_0}{e}\frac{\partial}{\partial z}\delta\mu(z,t) \label{spin-cur}
\end{equation}
where $\sigma_0$ is the electrical conductivity. 
Note that the generated spin current is not accompanied by charge current.
 By substituting the angular velocity Eq.~(\ref{z-component}) into Eq.~(\ref{spin-diff}) and solving the differential equation, we obtain
\begin{eqnarray}
\delta\mu(z,t) = 2\hbar \theta_0\omega_{\rm t}^2\int_{0}^{t}dt^\prime\int_{0}^{\infty}dz^\prime \frac{{\rm e}^{-\frac{t-t^\prime}{\tau_{\rm sf}}}}{\sqrt{4\pi D(t-t^\prime)}} \notag\\
\times\left[{\rm e}^{-\frac{(z^\prime-z)^2}{4D(t-t^\prime)}}+{\rm e}^{-\frac{(z^\prime+z)^2}{4D(t-t^\prime)}}\right] \sin{(q_{\rm t}z^\prime - \omega_{\rm t} t^\prime)} \label{variation} .
\end{eqnarray} 
By substituting Eq.~(\ref{variation}) into Eq.~(\ref{spin-cur}), we can numerically calculate the spin current. We set the realistic parameters of the CNT obtained in Ref.~\cite{parameter01}: $D = 8.0 \times 10^{-2} \ \mathrm{m^2/s}$ and  $\tau_{\rm sf} = 4.0 \times 10^{-11}\ \mathrm{s}$ at $0.25~\rm{K}$. 
As an example, we consider the $(10,10)$ armchair CNT with radius $R=6.78\ \mathrm{\AA}$, circumference $L=2\pi R=42.6\ \mathrm{\AA}$, the group velocity of the twisting mode $v_{\rm t} = 1.2\times 10^4\ \mathrm{m/s}$ \cite{Suzuura01}, and conductivity $\sigma_0 = 1.0 \times 10^6 ~\rm{(\Omega m)}^{-1}$ at $0.25~\rm{K}$ \cite{parameter02}. The twisting mode is assumed to be described by $q_{\rm t}=1.0\times 10^{6}\ \mathrm{m^{-1}}, \omega_{\rm t} = 1.2\times 10^{10}\ \mathrm{s^{-1}}$, and $\theta_0 = 0.2\pi$. 
 The result is shown in  Fig. \ref{spin-current}. The behavior can be classified into two regimes, $q_{\rm t}\lambda_{\rm sf} \lesssim 1$ (Fig.\ \ref{spin-current}\ (a)) and $q_{\rm t}\lambda_{\rm sf} > 1$ (Fig.\ \ref{spin-current}\ (b)) with spin diffusion length $\lambda_{\rm sf} = \sqrt{D\tau_{\rm sf}} 
\sim 1.8\times 10^{-6}\ \mathrm{m}$ 
. When $q_{\rm t}\lambda_{\rm sf} \lesssim 1$, the behavior of spin current is nearly the same as that of the twisting mode, since an influence of the reflection at the edge of CNT is negligible. This is because for $q_{\rm t} \lambda_{\rm sf} \lesssim 1$, the diffusion and relaxation  are much faster than the twisting mode. When $q_{\rm t}\lambda_{\rm sf} > 1$, the behavior changes around
the edge of CNT: the signal is strongly enhanced.  Sufficiently away from the edge, the behavior is nearly the same as that of the twisting mode and the magnitude of the spin current is proportional to the frequency of the mode. 
For $q_t \lambda_{\rm sf} > 1$, the spin current diffuses slowly and hence can be reflected at the edge.  


This theory requires presence of carriers, and therefore our theory applies to metallic CNT, doped semiconducting CNT, or semiconducting CNT at finite temperature. 
The resulting spin current depends on the chiralities and temperature through the physical quantities $\sigma_0$, $D$ and $\tau_{\rm sf}$. 
When the frequency and wavenumber are $\omega_{\rm t} = 1.2\times 10^{10}\ \mathrm{s^{-1}}$ and $ q_{\rm t} = 1.0\times 10^6 \ \mathrm{m^{-1}}$, the maximum of  spin current is evaluated as $J_S^z > 1.0 \times 10^6 \ \mathrm{A/m^{2}}$, as shown in Fig. \ref{spin-current} (b), which is measurable in experiments. 
By exciting the twisting mode in CNT with the piezoelectric device, the spin current can be measured via inverse spin Hall effect. For example, by making a contact between Pt and CNT, one can convert the spin current into a voltage signal. When a ferromagnet is attached to CNT, the generated spin current can be injected into the ferromagnet. 
This spin current can then exert spin torque on the magnetzation, leading to magnetization reversal.

{\it Orbital magnetization}. Next we investigate the orbital magnetization of CNTs by the orbital magnetic field. 
Here we focus on its radial component perpendicular to the tube axis. We follow the method to derive the orbital susceptibility for other magnetic field directions in Refs.~\cite{Ajiki01,Ajiki02}. For convenience in calculation, we introduce a flux $\phi_{\rm e}$ penetrating along the tube axis. We also assume that the frequency $\omega$ is sufficiently low and the response to the orbital magnetic field is regarded to be static. We use the cylindrical coordinate $(X,Y,Z)$ in Fig. \ref{structure}.

To calculate the orbital magnetization, we 
first calculate the shift of the energy levels up to the second order in the orbital magnetic field $\mathbf{B} =B\hat{\mathbf{Z}}$. The vector potential can be written as
$\mathbf{A}=BX\hat{\mathbf{Y}}$.
The $k \cdot p$ Hamiltonian near the $\rm K$ point is given by
\begin{equation}
\begin{pmatrix}
0 & \gamma(\hat{k}_X - i\hat{k}_Y)  \\
\gamma(\hat{k}_X + i\hat{k}_Y) & 0
\end{pmatrix}
\begin{pmatrix}
{F}_{\rm K}^{\rm A} \\
{F}_{\rm K}^{\rm B}
\end{pmatrix}
=
\varepsilon
\begin{pmatrix}
{F}_{\rm K}^{\rm A} \\
{F}_{\rm K}^{\rm B} \label{kpeq}
\end{pmatrix}
\end{equation}
with the envelope function near the $\rm K$ point ${F}_{\rm K}^{\rm A,B}$, the band parameter $\gamma \sim 6.45\  \mathrm{eV\AA} $, and $\mathbf{\hat{k}}$ being a wave vector operator defined by $\mathbf{\hat{k}} = -i\mathbf{\nabla} + e\mathbf{A}/\hbar$. In the absence of the orbital magnetic field along radial direction, $B=0$, the corresponding eigenvector is 
\begin{equation}
\mathbf{F}_{n}^s (k)= \frac{1}{\sqrt{2}}
\begin{pmatrix}
sb_{\nu\phi n}(k)\\
1
\end{pmatrix}, 
b_{\nu\phi n}(k) = \frac{\kappa_{\nu\phi n}-ik}{\sqrt{\kappa_{\nu\phi n}^2+k^2}}, 
\label{F}
\end{equation}
and the energy eigenvalue is $\varepsilon^\pm_{\nu\phi n} = \pm \sqrt{\kappa_{\nu\phi n}^2+k^2}$ where $\kappa_{\nu\phi n}= (2\pi/L)(n+\phi+\nu/3)$ is the wave number in the circumferential direction with circumference length $L$, $n$ is the band index, $\phi = \phi_{e}/\phi_0$ is a dimensionless parameter with the flux quantum $\phi_0 =h/e$, $\nu =0,\pm1$ correspond to metal and semiconductor CNT, respectively, $k$ is the wave number along the $Y$-direction, and $s=\pm1$ denotes the conduction and valence bands, respectively. 

\begin{figure}
\includegraphics[width=10cm]{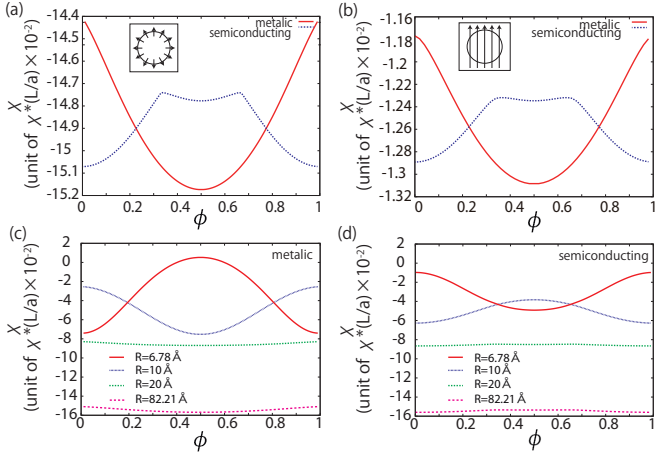}
\caption{\label{susceptibility}(Color online) (a), (b) Orbital susceptibility for the magnetic field in (a) radial direction and (b) direction perpendicular to the tube axis for the tube radius $R = 82.21\ \mathrm{\AA}$ \cite{Ajiki01}. Insets  show the direction of the orbiral magnetic field in the cross section of CNT. (c), (d) Orbital susceptibility for the magnetic field in radial direction for various values of the tube radius, for (c) metalic CNTs and (d) semiconducting CNTs.}
\end{figure}

The envelope function near the $K$ point and the vector potential
are then Fourier expanded 
in terms of $X$ 
in the range of $[-L/2,L/2]$.
Then, the perturbation due to the magnetic field $B$ is written as
\begin{equation}
H^{\prime} = -\frac{L\gamma}{2\pi l^2}\sum_{m=1}^{\infty}\frac{(-1)^m}{m}\left[{\rm e}^{i\frac{2\pi m}{L}X} - {\rm e}^{-i\frac{2\pi m}{L}X}\right]
\begin{pmatrix}
0 & -1\\
1 & 0
\end{pmatrix}
\end{equation}
with the magnetic length $l =\sqrt{\hbar/eB}$. The matrix elements of the perturbation 
$H'$ between the eigenstates for $B=0$ are calculated as,
\begin{equation}
\Braket{F_{n\pm m}^{s}|H^{\prime}|F_n^-} = \frac{\pm L\gamma}{4\pi l^2} 
\frac{(-1)^m}{m}[sb_{\nu\phi(n\pm m)}^*(k) + b_{\nu\phi n}(k)].
\end{equation}
Therefore, the magnetic field causes an energy shift for the eigenvalues $\varepsilon^{-}_{\nu\phi n}$ to the second order in $B$:
\begin{eqnarray}
\Delta\varepsilon_{\nu\phi n}^{-}(k) = \sum_{s=\pm1}\sum_{m=1}^{\infty}\frac{|\Braket{F_n^-|H^\prime|F_{n\pm m}^s}|^2}{\varepsilon_{\nu\phi n}^-(k)-\varepsilon_{\nu\phi(n\pm m)}^s(k)}.
\end{eqnarray}
At zero temperature, the shift of the total energy becomes
\begin{equation}
\Delta E=\frac{A}{2\pi}\sum_{n=-\infty}^{\infty}\int_{-\infty}^{\infty}dk\Delta \varepsilon_{\nu\phi n}^{-}(k)g(|\varepsilon_{\nu\phi n}^{-}(k)|)
\end{equation}
where $g(\varepsilon)$ is a cutoff function, and $A$ is the length of CNT. The susceptibility per unit area is calculated as $\chi = -\frac{1}{AL}(\partial^2\Delta E)/(\partial B^2)$.
Therefore, we obtain the susceptibility as
\begin{eqnarray}
\chi(\phi) &=& -\chi^*\frac{L}{a}\frac{2}{\pi^4}\sum_{n=-\infty}^{\infty}\sum_{m=1}^{\infty}\int_{-\infty}^{\infty}d\tilde{k}\frac{\sqrt{(n+\phi)^2+\tilde{k}^2}}{m^2(m^2-4(n+\phi)^2)}\notag \\
&\times& g(|\varepsilon_{\nu\phi n}^{-}(k)|) , \label{orbital sus}
\end{eqnarray}
with $\chi^* = \frac{2\pi \gamma}{a}\left(\frac{\pi a^2}{\phi_0}\right)^2\frac{1}{a^2}$, dimensionless parameter $\tilde{k}=Lk/2\pi$, and lattice constant $a = 2.46 \ \mathrm{\AA}$.
Adding the contribution from the $\rm K^\prime$ point and multiplying the result by a spin-degeneracy factor of two, 
we obtain the susceptibilities for metalic and semiconducting CNT ($\chi_M(\phi)$ and $\chi_S(\phi)$) as follows:
\begin{eqnarray}
\chi_M(\phi) &=&4\chi(\phi) \label{chiM}, \\
\chi_S(\phi) &=& 2\chi\left(\phi+\frac{1}{3}\right) + 2\chi\left(\phi - \frac{1}{3}\right).\label{chiS}
\end{eqnarray}
The result of the susceptibility for the radial magnetic field is shown in Fig. \ref{susceptibility}(a). For comparison, we show the susceptibility for a uniform magnetic field perpendicular to the tube axis
 in Fig. \ref{susceptibility}(b), using the method in Ref.~\cite{Ajiki01}. It is seen that the magnitude of the diamagnetic susceptibility for the radial magnetic field is about ten times larger than that for the magnetic field perpendicular to the tube axis. This is related with the enhanced diamagnetic response of the graphene sheet for the out-of-plane orbital 
magnetic field, when the Fermi energy is close to the Dirac point \cite{McClure}. 
We note that the susceptibility in a radial magnetic field survives above room temperature, as is similar to that in a magnetic field perpendicular to the tube axis in  Ref.~\cite{Tdependent}.

{\it Discussion}. We discuss the magnitude of the effective magnetic field. For 
 this purpose, we decompose the effective magnetic field
Eq.~(\ref{effective magnetic field}) into the component along the tube axis $B_{\rm a}$ and that along the radial direction $B_{\rm r}$:
\begin{align}
B_{\rm a} &= \frac{2m_{\rm e}}{e}\theta_0\omega_{\rm t}\cos{(q_{\rm t} z - \omega_{\rm t} t)}, \\
B_{\rm r} &= -\frac{2m_{\rm e}}{e}\theta_0 \omega_{\rm t}\frac{q_{\rm t} R}{2} \sin{(q_{\rm t} z - \omega_{\rm t} t)}.
\end{align}
We evaluate them for the $(10,10)$ armchair nanotube as an example. The twisting mode is characterized by the group velocity $v_{\rm t} = 12\ \mathrm{km/s}$, and the radius is  $R = 6.78\ \mathrm{\AA}$ \cite{Suzuura01}. For the frequency of the twisting mode $\omega_{\rm t}/2\pi=2\ \mathrm{GHz}$ and amplitude of the twisting mode $\theta_0 = 0.2\pi$, we have $q_{\rm t}R \simeq 6.8\times 10^{-4}$. 
The magnitude of the effective magnetic field is $B_{\rm a} \simeq 8.6 \times 10^{-2}\ \mathrm{T}$ and $B_{\rm r} \simeq -2.9 \times 10^{-5}\ \mathrm{T}$, and the ratio of $B_{\rm r}$ and $B_{\rm a}$ is 
\begin{equation}
\left|B_{\rm r}/B_{\rm a}\right| \sim q_{\rm t}R/2 =3.4\times 10^{-4}.
\end{equation}
The axial magnetic flux $\phi_{\rm a}$ is given by $\phi_{\rm a} = B_{\rm a}S$ where $S$ denotes the cross section of CNT. Hence, we have $\phi_{\rm a} =3.0\times 10^{-5}\phi_0$. Similarly, for the frequency $\omega_{\rm t}/2\pi= 200\ \mathrm{GHz}$, we have $q_{\rm t}R \simeq 6.8\times 10^{-2}$, $B_{\rm a} \simeq 8.6\ \mathrm{T}$, $B_{\rm r} \simeq -2.9\times10^{-1}\ \mathrm{T}$, and $\phi_{\rm a} \simeq 3.0\times10^{-3}\phi_0$. Therefore, when the frequency of the twisting mode is between $2\ \mathrm{GHz}$ and $200\ \mathrm{GHz}$, the ratio of $|B_r/B_a|$ is between $3.4\times 10^{-4}$ and $3.4\times 10^{-2}$, and the axial magnetic flux is  of the order from $ 10^{-5}\phi_0$ to $10^{-3}\phi_0$. Hence, the radially-polarized AC spin current  due to the Zeeman magnetic field is from $10^{-4}$ to $10^{-2}$ times smaller than that along the axial direction. Therefore, we can neglect the radial direction component $B_{\rm r}$, and the axial magnetic flux is very small.

Next, we roughly estimate four types of magnetizations: axial spin magnetization $M_{s\parallel}$, radial spin magnetization $M_{s\perp}$, axial orbital magnetization $M_{o\parallel}$, and radial orbital magnetization $M_{o\perp}$. The spin magnetization is given by the Pauli paramagnetic susceptibility $\chi_{\rm spin} = \mu_{\rm B}^2 D(\varepsilon_F)$ with density of state $D(\varepsilon)$. When the Fermi energy is $\varepsilon_{\rm F}= 0$, the density of state is $D(0) = 4/\pi\gamma$. Then the magnetization per unit area is $M_{\rm s\parallel(\perp)} = \chi_{\rm spin }B_{{\rm a}({\rm r})}/L$.

The orbital susceptibility $\chi_{\rm orbital}$ by the radial magnetic field is given by Eqs.~(\ref{chiM}) and (\ref{chiS}), and is shown in Figs. \ref{susceptibility}(c) and (d) for various values of the radii. It can be seen that for thicker CNT, the susceptibility becomes larger and dependence on $\phi$ becomes weaker. These behaviors are similar to those in response to uniform magnetic field perpendicular to the tube~\cite{Ajiki01}. 
For a rough estimate of the magnetization, 
we use the averaged value of the susceptibility from $\phi=0$ to $\phi=1$, $\chi_{\rm orbital}=-3.1\chi^*(L/a)\times10^{-2}$, which is common for metallic and semiconducting CNTs. The magnetization per unit area is $M_{\rm o\perp} = \chi_{\rm orbital} B_{{\rm r}}$. For the axial orbital magnetization $M_{o\parallel}$, the susceptibility diverges at $\phi=0$ \cite{Ajiki01}, and therefore the response is nonlinear in $B_a$. Therefore, we use the results in Ref.~\cite{Ajiki01} to calculate  $M_{o\parallel}$ for 
the given value of $B_a$

In the frequency range from $\omega_{\rm t}/2\pi= 2 \ \mathrm{GHz}$ to $200\ \mathrm{GHz}$, the magnitudes of the four magnetizations are evaluated as follows:  $M_{s\parallel}=2.1\times 10^{-11}$ to $2.1\times 10^{-9}\ \mathrm{J/Tm^2}$, $M_{s\perp}=-7.3\times 10^{-15}$ to $-7.3\times 10^{-11}\ \mathrm{J/Tm^2}$, $M_{o\parallel}=5.0\times 10^{-9}$ to $3.5\times 10^{-7}\ \mathrm{J/Tm^2}$, and $M_{o\perp}=1.4\times 10^{-12}$ to $1.4\times 10^{-8}\ \mathrm{J/Tm^2}$. We find that the orbital magnetizations $M_{o}$ are larger than the spin magnetizations $M_{s}$ in both directions, and axial magnetizations $M_{\parallel}$ are larger than radial magnetizations $M_{\perp}$. The axial orbital magnetization $M_{o\parallel}$ is the largest among the four
 in this setup, and is the main contribution. 
The magnetization can be measured by the magnet-optic Kerr effect or using SQUID.
 {\it Summary}. In this paper,
we have theoretically investigated spin current and magnetic response induced by the twisting phonon mode in carbon nanotubes via the spin-rotation coupling. 
We found that the AC spin current and orbital magnetizations are generated by the twisting phonon mode. The spin current is detectable in magnitude when the frequency is of the order of GHz. It was shown that the orbital magnetizations are larger than the spin magnetizations, and the axial orbital magnetization is the main contribution for realistic values of the parameters.
\begin{acknowledgments}
We thank M. Matsuo, J. Ieda, K. Harii, and S. Maekawa for useful discussions.
This work was supported by Grant-in-Aid for Young Scientists (B) (No. 23740236) , the ``Topological Quantum Phenomena" (No. 25103709) Grant-in-Aid for Scientific Research on Innovative Areas (No.26103006), 
 and MEXT Elements Strategy Initiative to Form Core Research Center (TIES) from the Ministry of Education, Culture, Sports, Science and Technology (MEXT) of Japan.

\end{acknowledgments}

\begin{thebibliography}{17}%
\makeatletter
\providecommand \@ifxundefined [1]{%
 \@ifx{#1\undefined}
}%
\providecommand \@ifnum [1]{%
 \ifnum #1\expandafter \@firstoftwo
 \else \expandafter \@secondoftwo
 \fi
}%
\providecommand \@ifx [1]{%
 \ifx #1\expandafter \@firstoftwo
 \else \expandafter \@secondoftwo
 \fi
}%
\providecommand \natexlab [1]{#1}%
\providecommand \enquote  [1]{``#1''}%
\providecommand \bibnamefont  [1]{#1}%
\providecommand \bibfnamefont [1]{#1}%
\providecommand \citenamefont [1]{#1}%
\providecommand \href@noop [0]{\@secondoftwo}%
\providecommand \href [0]{\begingroup \@sanitize@url \@href}%
\providecommand \@href[1]{\@@startlink{#1}\@@href}%
\providecommand \@@href[1]{\endgroup#1\@@endlink}%
\providecommand \@sanitize@url [0]{\catcode `\\12\catcode `\$12\catcode
  `\&12\catcode `\#12\catcode `\^12\catcode `\_12\catcode `\%12\relax}%
\providecommand \@@startlink[1]{}%
\providecommand \@@endlink[0]{}%
\providecommand \url  [0]{\begingroup\@sanitize@url \@url }%
\providecommand \@url [1]{\endgroup\@href {#1}{\urlprefix }}%
\providecommand \urlprefix  [0]{URL }%
\providecommand \Eprint [0]{\href }%
\providecommand \doibase [0]{http://dx.doi.org/}%
\providecommand \selectlanguage [0]{\@gobble}%
\providecommand \bibinfo  [0]{\@secondoftwo}%
\providecommand \bibfield  [0]{\@secondoftwo}%
\providecommand \translation [1]{[#1]}%
\providecommand \BibitemOpen [0]{}%
\providecommand \bibitemStop [0]{}%
\providecommand \bibitemNoStop [0]{.\EOS\space}%
\providecommand \EOS [0]{\spacefactor3000\relax}%
\providecommand \BibitemShut  [1]{\csname bibitem#1\endcsname}%
\let\auto@bib@innerbib\@empty
\bibitem [{\citenamefont {Kane}\ \emph {et~al.}(1998)\citenamefont {Kane},
  \citenamefont {Mele}, \citenamefont {Lee}, \citenamefont {Fischer},
  \citenamefont {Petit}, \citenamefont {Dai}, \citenamefont {Thess},
  \citenamefont {Smalley}, \citenamefont {Verschueren}, \citenamefont {Tans},\
  and\ \citenamefont {Dekker}}]{Kane1}%
  \BibitemOpen
  \bibfield  {author} {\bibinfo {author} {\bibfnamefont {C.~L.}\ \bibnamefont
  {Kane}}, \bibinfo {author} {\bibfnamefont {E.~J.}\ \bibnamefont {Mele}},
  \bibinfo {author} {\bibfnamefont {R.~S.}\ \bibnamefont {Lee}}, \bibinfo
  {author} {\bibfnamefont {J.~E.}\ \bibnamefont {Fischer}}, \bibinfo {author}
  {\bibfnamefont {P.}~\bibnamefont {Petit}}, \bibinfo {author} {\bibfnamefont
  {H.}~\bibnamefont {Dai}}, \bibinfo {author} {\bibfnamefont {A.}~\bibnamefont
  {Thess}}, \bibinfo {author} {\bibfnamefont {R.~E.}\ \bibnamefont {Smalley}},
  \bibinfo {author} {\bibfnamefont {A.~R.~M.}\ \bibnamefont {Verschueren}},
  \bibinfo {author} {\bibfnamefont {S.~J.}\ \bibnamefont {Tans}}, \ and\
  \bibinfo {author} {\bibfnamefont {C.}~\bibnamefont {Dekker}},\ }\href
  {http://stacks.iop.org/0295-5075/41/i=6/a=683} {\bibfield  {journal}
  {\bibinfo  {journal} {EPL (Europhysics Letters)}\ }\textbf {\bibinfo {volume}
  {41}},\ \bibinfo {pages} {683} (\bibinfo {year} {1998})}\BibitemShut
  {NoStop}%
\bibitem [{\citenamefont {Kane}\ and\ \citenamefont {Mele}(1997)}]{TW01}%
  \BibitemOpen
  \bibfield  {author} {\bibinfo {author} {\bibfnamefont {C.~L.}\ \bibnamefont
  {Kane}}\ and\ \bibinfo {author} {\bibfnamefont {E.~J.}\ \bibnamefont
  {Mele}},\ }\href {\doibase 10.1103/PhysRevLett.78.1932} {\bibfield  {journal}
  {\bibinfo  {journal} {Phys. Rev. Lett.}\ }\textbf {\bibinfo {volume} {78}},\
  \bibinfo {pages} {1932} (\bibinfo {year} {1997})}\BibitemShut {NoStop}%
\bibitem [{\citenamefont {S\'anchez-Portal}\ \emph {et~al.}(1999)\citenamefont
  {S\'anchez-Portal}, \citenamefont {Artacho}, \citenamefont {Soler},
  \citenamefont {Rubio},\ and\ \citenamefont {Ordej\'on}}]{TW02}%
  \BibitemOpen
  \bibfield  {author} {\bibinfo {author} {\bibfnamefont {D.}~\bibnamefont
  {S\'anchez-Portal}}, \bibinfo {author} {\bibfnamefont {E.}~\bibnamefont
  {Artacho}}, \bibinfo {author} {\bibfnamefont {J.~M.}\ \bibnamefont {Soler}},
  \bibinfo {author} {\bibfnamefont {A.}~\bibnamefont {Rubio}}, \ and\ \bibinfo
  {author} {\bibfnamefont {P.}~\bibnamefont {Ordej\'on}},\ }\href {\doibase
  10.1103/PhysRevB.59.12678} {\bibfield  {journal} {\bibinfo  {journal} {Phys.
  Rev. B}\ }\textbf {\bibinfo {volume} {59}},\ \bibinfo {pages} {12678}
  (\bibinfo {year} {1999})}\BibitemShut {NoStop}%
\bibitem [{\citenamefont {Matsuo}\ \emph {et~al.}(2011)\citenamefont {Matsuo},
  \citenamefont {Ieda}, \citenamefont {Saitoh},\ and\ \citenamefont
  {Maekawa}}]{Matsuo02}%
  \BibitemOpen
  \bibfield  {author} {\bibinfo {author} {\bibfnamefont {M.}~\bibnamefont
  {Matsuo}}, \bibinfo {author} {\bibfnamefont {J.}~\bibnamefont {Ieda}},
  \bibinfo {author} {\bibfnamefont {E.}~\bibnamefont {Saitoh}}, \ and\ \bibinfo
  {author} {\bibfnamefont {S.}~\bibnamefont {Maekawa}},\ }\href {\doibase
  10.1103/PhysRevLett.106.076601} {\bibfield  {journal} {\bibinfo  {journal}
  {Phys. Rev. Lett.}\ }\textbf {\bibinfo {volume} {106}},\ \bibinfo {pages}
  {076601} (\bibinfo {year} {2011})}\BibitemShut {NoStop}%
\bibitem [{\citenamefont {de~Oliveira}\ and\ \citenamefont
  {Tiomno}(1962)}]{SRC01}%
  \BibitemOpen
  \bibfield  {author} {\bibinfo {author} {\bibfnamefont {C.}~\bibnamefont
  {de~Oliveira}}\ and\ \bibinfo {author} {\bibfnamefont {J.}~\bibnamefont
  {Tiomno}},\ }\href {\doibase 10.1007/BF02816716} {\bibfield  {journal}
  {\bibinfo  {journal} {Il Nuovo Cimento}\ }\textbf {\bibinfo {volume} {24}},\
  \bibinfo {pages} {672} (\bibinfo {year} {1962})}\BibitemShut {NoStop}%
\bibitem [{\citenamefont {Mashhoon}(1988)}]{SRC02}%
  \BibitemOpen
  \bibfield  {author} {\bibinfo {author} {\bibfnamefont {B.}~\bibnamefont
  {Mashhoon}},\ }\href {\doibase 10.1103/PhysRevLett.61.2639} {\bibfield
  {journal} {\bibinfo  {journal} {Phys. Rev. Lett.}\ }\textbf {\bibinfo
  {volume} {61}},\ \bibinfo {pages} {2639} (\bibinfo {year}
  {1988})}\BibitemShut {NoStop}%
\bibitem [{\citenamefont {Matsuo}\ \emph
  {et~al.}(2013{\natexlab{a}})\citenamefont {Matsuo}, \citenamefont {Ieda},
  \citenamefont {Harii}, \citenamefont {Saitoh},\ and\ \citenamefont
  {Maekawa}}]{Matsuo01}%
  \BibitemOpen
  \bibfield  {author} {\bibinfo {author} {\bibfnamefont {M.}~\bibnamefont
  {Matsuo}}, \bibinfo {author} {\bibfnamefont {J.}~\bibnamefont {Ieda}},
  \bibinfo {author} {\bibfnamefont {K.}~\bibnamefont {Harii}}, \bibinfo
  {author} {\bibfnamefont {E.}~\bibnamefont {Saitoh}}, \ and\ \bibinfo {author}
  {\bibfnamefont {S.}~\bibnamefont {Maekawa}},\ }\href {\doibase
  10.1103/PhysRevB.87.180402} {\bibfield  {journal} {\bibinfo  {journal} {Phys.
  Rev. B}\ }\textbf {\bibinfo {volume} {87}},\ \bibinfo {pages} {180402}
  (\bibinfo {year} {2013}{\natexlab{a}})}\BibitemShut {NoStop}%
\bibitem [{\citenamefont {Barnett}(1915)}]{Barnet}%
  \BibitemOpen
  \bibfield  {author} {\bibinfo {author} {\bibfnamefont {S.~J.}\ \bibnamefont
  {Barnett}},\ }\href {\doibase 10.1103/PhysRev.6.239} {\bibfield  {journal}
  {\bibinfo  {journal} {Phys. Rev.}\ }\textbf {\bibinfo {volume} {6}},\
  \bibinfo {pages} {239} (\bibinfo {year} {1915})}\BibitemShut {NoStop}%
\bibitem [{\citenamefont {Tombros}\ \emph {et~al.}(2006)\citenamefont
  {Tombros}, \citenamefont {van~der Molen},\ and\ \citenamefont {van
  Wees}}]{experiment1}%
  \BibitemOpen
  \bibfield  {author} {\bibinfo {author} {\bibfnamefont {N.}~\bibnamefont
  {Tombros}}, \bibinfo {author} {\bibfnamefont {S.~J.}\ \bibnamefont {van~der
  Molen}}, \ and\ \bibinfo {author} {\bibfnamefont {B.~J.}\ \bibnamefont {van
  Wees}},\ }\href {\doibase 10.1103/PhysRevB.73.233403} {\bibfield  {journal}
  {\bibinfo  {journal} {Phys. Rev. B}\ }\textbf {\bibinfo {volume} {73}},\
  \bibinfo {pages} {233403} (\bibinfo {year} {2006})}\BibitemShut {NoStop}%
\bibitem [{\citenamefont {Yang}\ \emph {et~al.}(2012)\citenamefont {Yang},
  \citenamefont {Itkis}, \citenamefont {Moriya}, \citenamefont {Rettner},
  \citenamefont {Jeong}, \citenamefont {Pickard}, \citenamefont {Haddon},\ and\
  \citenamefont {Parkin}}]{parameter01}%
  \BibitemOpen
  \bibfield  {author} {\bibinfo {author} {\bibfnamefont {H.}~\bibnamefont
  {Yang}}, \bibinfo {author} {\bibfnamefont {M.~E.}\ \bibnamefont {Itkis}},
  \bibinfo {author} {\bibfnamefont {R.}~\bibnamefont {Moriya}}, \bibinfo
  {author} {\bibfnamefont {C.}~\bibnamefont {Rettner}}, \bibinfo {author}
  {\bibfnamefont {J.-S.}\ \bibnamefont {Jeong}}, \bibinfo {author}
  {\bibfnamefont {D.~S.}\ \bibnamefont {Pickard}}, \bibinfo {author}
  {\bibfnamefont {R.~C.}\ \bibnamefont {Haddon}}, \ and\ \bibinfo {author}
  {\bibfnamefont {S.~S.~P.}\ \bibnamefont {Parkin}},\ }\href {\doibase
  10.1103/PhysRevB.85.052401} {\bibfield  {journal} {\bibinfo  {journal} {Phys.
  Rev. B}\ }\textbf {\bibinfo {volume} {85}},\ \bibinfo {pages} {052401}
  (\bibinfo {year} {2012})}\BibitemShut {NoStop}%
\bibitem [{\citenamefont {Suzuura}\ and\ \citenamefont
  {Ando}(2002)}]{Suzuura01}%
  \BibitemOpen
  \bibfield  {author} {\bibinfo {author} {\bibfnamefont {H.}~\bibnamefont
  {Suzuura}}\ and\ \bibinfo {author} {\bibfnamefont {T.}~\bibnamefont {Ando}},\
  }\href {\doibase 10.1103/PhysRevB.65.235412} {\bibfield  {journal} {\bibinfo
  {journal} {Phys. Rev. B}\ }\textbf {\bibinfo {volume} {65}},\ \bibinfo
  {pages} {235412} (\bibinfo {year} {2002})}\BibitemShut {NoStop}%
\bibitem [{\citenamefont {Matsuo}\ \emph
  {et~al.}(2013{\natexlab{b}})\citenamefont {Matsuo}, \citenamefont {Ieda},\
  and\ \citenamefont {Maekawa}}]{Matsuo03}%
  \BibitemOpen
  \bibfield  {author} {\bibinfo {author} {\bibfnamefont {M.}~\bibnamefont
  {Matsuo}}, \bibinfo {author} {\bibfnamefont {J.}~\bibnamefont {Ieda}}, \ and\
  \bibinfo {author} {\bibfnamefont {S.}~\bibnamefont {Maekawa}},\ }\href
  {\doibase 10.1103/PhysRevB.87.115301} {\bibfield  {journal} {\bibinfo
  {journal} {Phys. Rev. B}\ }\textbf {\bibinfo {volume} {87}},\ \bibinfo
  {pages} {115301} (\bibinfo {year} {2013}{\natexlab{b}})}\BibitemShut
  {NoStop}%
\bibitem [{\citenamefont {Fischer}\ \emph {et~al.}(1997)\citenamefont
  {Fischer}, \citenamefont {Dai}, \citenamefont {Thess}, \citenamefont {Lee},
  \citenamefont {Hanjani}, \citenamefont {Dehaas},\ and\ \citenamefont
  {Smalley}}]{parameter02}%
  \BibitemOpen
  \bibfield  {author} {\bibinfo {author} {\bibfnamefont {J.~E.}\ \bibnamefont
  {Fischer}}, \bibinfo {author} {\bibfnamefont {H.}~\bibnamefont {Dai}},
  \bibinfo {author} {\bibfnamefont {A.}~\bibnamefont {Thess}}, \bibinfo
  {author} {\bibfnamefont {R.}~\bibnamefont {Lee}}, \bibinfo {author}
  {\bibfnamefont {N.~M.}\ \bibnamefont {Hanjani}}, \bibinfo {author}
  {\bibfnamefont {D.~L.}\ \bibnamefont {Dehaas}}, \ and\ \bibinfo {author}
  {\bibfnamefont {R.~E.}\ \bibnamefont {Smalley}},\ }\href {\doibase
  10.1103/PhysRevB.55.R4921} {\bibfield  {journal} {\bibinfo  {journal} {Phys.
  Rev. B}\ }\textbf {\bibinfo {volume} {55}},\ \bibinfo {pages} {R4921}
  (\bibinfo {year} {1997})}\BibitemShut {NoStop}%
\bibitem [{\citenamefont {Ajiki}\ and\ \citenamefont {Ando}(1993)}]{Ajiki01}%
  \BibitemOpen
  \bibfield  {author} {\bibinfo {author} {\bibfnamefont {H.}~\bibnamefont
  {Ajiki}}\ and\ \bibinfo {author} {\bibfnamefont {T.}~\bibnamefont {Ando}},\
  }\href@noop {} {\bibfield  {journal} {\bibinfo  {journal} {J. Phys. Soc.
  Jpn.}\ }\textbf {\bibinfo {volume} {62}},\ \bibinfo {pages} {2470} (\bibinfo
  {year} {1993})}\BibitemShut {NoStop}%
\bibitem [{\citenamefont {Ajiki}\ and\ \citenamefont {Ando}(1992)}]{Ajiki02}%
  \BibitemOpen
  \bibfield  {author} {\bibinfo {author} {\bibfnamefont {H.}~\bibnamefont
  {Ajiki}}\ and\ \bibinfo {author} {\bibfnamefont {T.}~\bibnamefont {Ando}},\
  }\href@noop {} {\bibfield  {journal} {\bibinfo  {journal} {J. Phys. Soc.
  Jpn.}\ }\textbf {\bibinfo {volume} {62}},\ \bibinfo {pages} {1255} (\bibinfo
  {year} {1992})}\BibitemShut {NoStop}%
\bibitem [{\citenamefont {McClure}(1956)}]{McClure}%
  \BibitemOpen
  \bibfield  {author} {\bibinfo {author} {\bibfnamefont {J.~W.}\ \bibnamefont
  {McClure}},\ }\href {\doibase 10.1103/PhysRev.104.666} {\bibfield  {journal}
  {\bibinfo  {journal} {Phys. Rev.}\ }\textbf {\bibinfo {volume} {104}},\
  \bibinfo {pages} {666} (\bibinfo {year} {1956})}\BibitemShut {NoStop}%
\bibitem [{\citenamefont {Lu}(1995)}]{Tdependent}%
  \BibitemOpen
  \bibfield  {author} {\bibinfo {author} {\bibfnamefont {J.~P.}\ \bibnamefont
  {Lu}},\ }\href {\doibase 10.1103/PhysRevLett.74.1123} {\bibfield  {journal}
  {\bibinfo  {journal} {Phys. Rev. Lett.}\ }\textbf {\bibinfo {volume} {74}},\
  \bibinfo {pages} {1123} (\bibinfo {year} {1995})}\BibitemShut {NoStop}%
\end{thebibliography}
%

\end{document}